\documentclass[reprint,
amsmath,amssymb,11pt,reprint,aps,prl,showpacs,floatfix]{revtex4-2}

\usepackage[english]{babel}
\usepackage{multirow}
\usepackage{xspace}
\usepackage{ulem}
\usepackage{soul}
\usepackage{graphicx}% Include figure files
\graphicspath{{figures//}} %allows to omit the path in the \incudegraphics command
\usepackage{dcolumn}% Align table columns on decimal point
\usepackage{bm}% bold math
\usepackage{color}
\usepackage{wrapfig}
\usepackage{hyperref}
\usepackage{array}
\usepackage{booktabs}
\usepackage{multirow}
\usepackage{amssymb}

\usepackage{marvosym}
\hypersetup{
    unicode=false,          % non-Latin characters in Acrobat’s bookmarks
    pdftoolbar=true,        % show Acrobat’s toolbar?
    pdfmenubar=true,        % show Acrobat’s menu?
    pdffitwindow=false,     % window fit to page when opened
    pdfstartview={FitH},    % fits the width of the page to the window
    pdftitle={My title},    % title
    pdfauthor={Author},     % author
    pdfsubject={Subject},   % subject of the document
    pdfcreator={Creator},   % creator of the document
    pdfproducer={Producer}, % producer of the document
    pdfkeywords={keyword1} {key2} {key3}, % list of keywords
    pdfnewwindow=true,      % links in new window
    colorlinks=true,       % false: boxed links; true: colored links
    linkcolor=blue,          % color of internal links (change box color with linkbordercolor)
    citecolor=blue,        % color of links to bibliography
    filecolor=blue,      % color of file links
    urlcolor=blue,       % color of external links
    plainpages=false        % influences the page numbering
}

\newcommand{\TSDW}{$T_{\text{SDW}}$\xspace}

\newcommand{\LNO}{\mbox{La$_3$Ni$_2$O$_7$}\,\xspace}

\newcommand{\Blg}{${B_{\rm{1g}}}$\xspace}

\newcommand{\wn}{\rm{cm}$^{-1}$\,}

\begin{document}

\title{ \Large Unconventional Pressure Evolution of Spin-Density-Wave State in \LNO}

\author{Xiaoxiang Zhou$^{1,7}$,
Shiyu Xie$^{2,7}$, 
Liangxin Qiao$^{2,7}$,
Hengyuan Zhang$^{3,7}$,
Jun Shu$^{4,5}$,
Rui Liu$^{1}$,
Mengwu Huo$^{3}$,
Deyuan Hu$^{3}$,
Hengjie Liu$^{2}$, 
Chuansheng Hu$^{2}$, 
Ge He$^{4,6}$,
Zeming Qi$^{2}$\textrm{\Letter},
Meng Wang$^{3}$\textrm{\Letter},
Dong-Lai Feng$^{1}$\textrm{\Letter},
Zengyi Du$^{1}$\textrm{\Letter}}
\affiliation{
$^1$ Hefei National Laboratory\mbox{,} and New Cornerstone Science Laboratory\mbox{,} Hefei, Anhui 230088\mbox{,} China\\ 
$^2$ National Synchrotron Radiation Laboratory\mbox{,} University of Science and Technology of China\mbox{,} Hefei\mbox{,} Anhui 230029\mbox{,} China\\
$^3$ Center for Neutron Science and Technology\mbox{,} Guangdong Provincial Key Laboratory of Magnetoelectric Physics and Devices\mbox{,} School of Physics\mbox{,} Sun Yat-Sen University\mbox{,} Guangzhou 510275\mbox{,} China\\
$^4$ School of Mechanical Engineering\mbox{,} Beijing Institute of Technology\mbox{,} Beijing 100081\mbox{,} China \\
$^5$ Department of Applied Physics\mbox{,} Wuhan University of Science and Technology\mbox{,} Wuhan 430081\mbox{,} China\\
$^6$ Beijing Key Laboratory of Quantum Matter State Control and Ultra-Precision Measurement Technology\mbox{,} Beijing Institute of Technology\mbox{,} Beijing 100081\mbox{,} China\\
$^{7}$ These authors contributed equally: Xiaoxiang Zhou, Shiyu Xie, Liangxin Qiao and Hengyuan Zhang\\
\textrm{\Letter} e-mail: 
zmqi@ustc.edu.cn
wangmeng5@mail.sysu.edu.cn
dlfeng@ustc.edu.cn
duzengyi@hfnl.cn
}

\date{\today}
\begin{abstract}
The discovery of pressure-induced high temperature superconductivity in the bilayer nickelate \LNO\ has raised the question of how its spin-density-wave (SDW) state evolves toward the superconducting regime. Here, we report a systematic electronic Raman study of \LNO single crystals under hydrostatic pressures up to 16.51~GPa. Both the SDW gap energy and the transition temperature $T_{\mathrm{SDW}}$ show an overall increase with pressure, while the dimensionless coupling ratio $2\Delta_{\mathrm{SDW}}/(k_{\mathrm B}T_{\mathrm{SDW}})$ remains approximately pressure independent, indicating a robust strong-coupling character of SDW state. At the same time, the SDW-related Raman peak broadens as pressure is applied, indicating the enhanced damping of the associated electronic excitation. These results reveal an unusual pressure evolution in which the SDW energy scale is enhanced while the SDW state becomes progressively less coherent, providing spectroscopic constraints on the magnetic correlations relevant to superconductivity in bilayer nickelates.

\end{abstract}

%\pacs{74.50.+r, 73.23.Ad, 74.45.+c, 73.40.Gk}

\maketitle

\section{Introduction}
High-temperature superconductivity commonly emerges in close proximity to antiferromagnetic or spin-density-wave (SDW) order in the phase diagram~\cite{Keimer:2015,Dai:2015}, raising the fundamental question of how magnetic order evolves as superconductivity is approached. The pressurized nickelates exhibit high-$T_{\mathrm{c}}$ superconductivity~\cite{Wang:2024,Sun:2023,Zhang:2024,WangbulkSC:2024,Zhu:2024,Zhangjunjie:2025}, providing a new platform for investigating this interplay.

The low-energy physics of \LNO\ has been described within a bilayer $t$\text{\ensuremath{-}}$J$ model featuring strong interlayer exchange interactions~\cite{Wucongjun:2024,Sugang:2024}. This strong-coupling picture is supported by experimental observations of pronounced band renormalization and spectral-weight transfer~\cite{Zhouxingjiang:2024,Liuzhe:2024,GeHe:2025}. Furthermore, resonant inelastic x-ray scattering (RIXS), inelastic neutron scattering, and transport measurements have revealed substantial interlayer spin correlations in this system~\cite{RIXS:2024,Zhou2026SCPMA,Liu2026arXiv.2605.18524}. Within this framework, pressure is expected to enhance this interlayer exchange coupling and modify the characteristic magnetic energy scales~\cite{Sun:2023,Wucongjun:2024,Sugang:2024,Liu2026arXiv.2605.18524,Verraes2025arXiv.2502.19501,IlyaCDWPRB:2026}. Experimentally, the pressure evolution of magnetic ordering has been partially established: nuclear magnetic resonance (NMR) and muon spin rotation ($\mu$SR) measurements show that the SDW transition temperature $T_{\mathrm{SDW}}$ increases monotonically with pressure~\cite{Zhao2025SB.70.1239,Khasanov:2025}. However, these measurements are limited to pressures below approximately 3~GPa, far below the superconducting regime emerging near 14~GPa~\cite{Sun:2023}. Transport measurements on the intergrowth compound $\mathrm{La_5Ni_3O_{11}}$ suggest that $T_{\mathrm{SDW}}$ continues to increase up to approximately 13~GPa~\cite{Shi2025NP.21.1780}. Nevertheless, existing high-pressure studies have primarily established the evolution of phase boundaries and transition temperatures~\cite{Hou2023CPL.40.117302,Zhao2025SB.70.1239,Khasanov:2025,Shi2025NP.21.1780,Yuxiaohui:2024,Kriener:2026,HengyuanZhang:2025,FrancescoCapitani:2026}, while the pressure evolution of the SDW-related electronic excitation spectrum in \LNO remains poorly understood. It therefore remains unclear how the characteristic SDW energy scale evolves upon entering the superconducting pressure regime.

%%%%%%%%%%%%%%%%%%%%%%%%%%%%%%%%%%%%%%%%%%%%%
Electronic Raman scattering provides a symmetry-resolved probe of low-energy electronic excitations at nearly zero momentum transfer. Through the polarization dependence of the Raman vertex, different symmetry channels selectively sample electronic excitations from different regions of momentum space, making Raman spectroscopy particularly sensitive to changes in the electronic continuum associated with symmetry breaking, gap opening, and electronic reconstruction~\cite{Devereaux:2007,GeHe:2025,Kretzschmar:2016,Bohm2014PRX}. Previous electronic Raman scattering studies have consistently reported the opening of a density-wave (DW) gap accompanied by a transfer of spectral weight from low to high energies~\cite{Chauviere2010PRB.82.180521,Chauviere2011PRB.84.104508,Loret2019NP.15.771,Suthar2025arXiv,Gim2026arXiv.2602.05365,GeHe:2025,Sundaramurthy2025arXiv}. Polarization-resolved electronic Raman measurements on \LNO further revealed a clear redistribution of spectral weight in the $B_{1g}$ and $B_{2g}$ channels below $T_{\mathrm{SDW}}\approx150$~K, which was attributed to the opening of anisotropic SDW gaps~\cite{GeHe:2025}. By contrast, Gim \textit{et al}. reported similar Raman responses but argued that the prominent $B_{1g}$ and $B_{2g}$ features are dominated by two-magnon excitations associated with magnetic exchange processes rather than DW-gap excitations~\cite{Gim2026arXiv.2602.05365}. Although different studies have proposed distinct microscopic interpretations, both associate the observed Raman features with the magnetic state of \LNO at ambient pressure. Accordingly, throughout this work we associate the broad Raman peak with the SDW state based on its temperature evolution and its correspondence with the SDW phase established by magnetic probes, while emphasizing that the Raman response reflects the accompanying electronic reconstruction rather than the finite-$q$ magnetic order directly. How these symmetry-resolved Raman features evolve under pressure, however, remains an open question.

In this study, we investigate the pressure evolution of the SDW-related Raman response under hydrostatic pressure up to 16.51~GPa. While SDW-related Raman features have been reported in both the $B_{1g}$ and $B_{2g}$ channels at ambient pressure~\cite{GeHe:2025}, no clearly distinguishable SDW-related Raman peak is resolved in the $B_{2g}$ channel under pressure in the present measurements. By contrast, the $B_{1g}$ SDW-related Raman peak remains well defined over a broad pressure range, allowing its characteristic energy and linewidth to be systematically tracked. We therefore focus the following pressure-dependent analysis on the $B_{1g}$ channel. Both the SDW gap energy and $T_{\mathrm{SDW}}$ increase monotonically with pressure, revealing an enhanced SDW energy scale. Meanwhile, the dimensionless coupling ratio $2\Delta_{\mathrm{SDW}}/(k_{\mathrm B}T_{\mathrm{SDW}})$ remains approximately constant, indicating that the strong-coupling character of the SDW state is preserved under pressure. Simultaneously, the SDW-related Raman peak progressively broadens, suggesting enhanced damping and reduced spectral coherence of the associated electronic excitation under pressure. This work provides the first Raman spectroscopic characterization of the pressure evolution of the SDW state in bilayer nickelates, offering new insight into the magnetic correlations associated with high-temperature superconductivity.

\section{Results}
High-quality \LNO\ single crystals were grown by the vertical optical floating-zone method under an oxygen pressure of 15~bar using a 5~kW xenon arc lamp. A detailed characterization of the crystal quality is provided in Supplemental Material A of Ref.~\cite{GeHe:2025}. Crystals with shiny and flat surfaces were selected for Raman measurements. High-pressure and low-temperature polarization-resolved Raman measurements were carried out at the Infrared Spectroscopy and Microspectroscopy Beamline of the Hefei Light Source using a home-built Raman system integrated with a membrane-driven diamond anvil cell (DAC) and a He-flow cryostat. The Raman spectra were collected with a Princeton Instruments SpectraPro HRS-500 spectrometer equipped with a 1200~grooves/mm grating and a PIXIS 256E CCD detector. A 532~nm diode-pumped solid-state laser was used for excitation, with the laser power maintained at approximately 1.3~mW at the sample surface. The laser beam was focused through a $20\times$ objective to a spot approximately 10~$\mu$m in diameter. High pressure was generated using a membrane-driven DAC with argon as the pressure-transmitting medium. Type-IIa diamond anvils with a culet diameter of 550~$\mu$m were used, and the pressure was calibrated by ruby fluorescence. For temperature-dependent measurements, the DAC was mounted on the cold finger of a He-flow cryostat (JANIS ST-100). At each pressure, the sample was first cooled to 20~K, and Raman spectra were subsequently collected upon warming to 300~K. The temperature was controlled by regulating the liquid-helium flow together with resistive heating. This procedure was repeated at each investigated pressure, with the pressure tuned from ambient pressure to 16.51~GPa by adjusting the gas pressure applied to the DAC membrane (see Supplemental Material~\cite{supplement} for more details).

\begin{figure}[ht!]
  \centering
  \includegraphics[width=8.5cm]{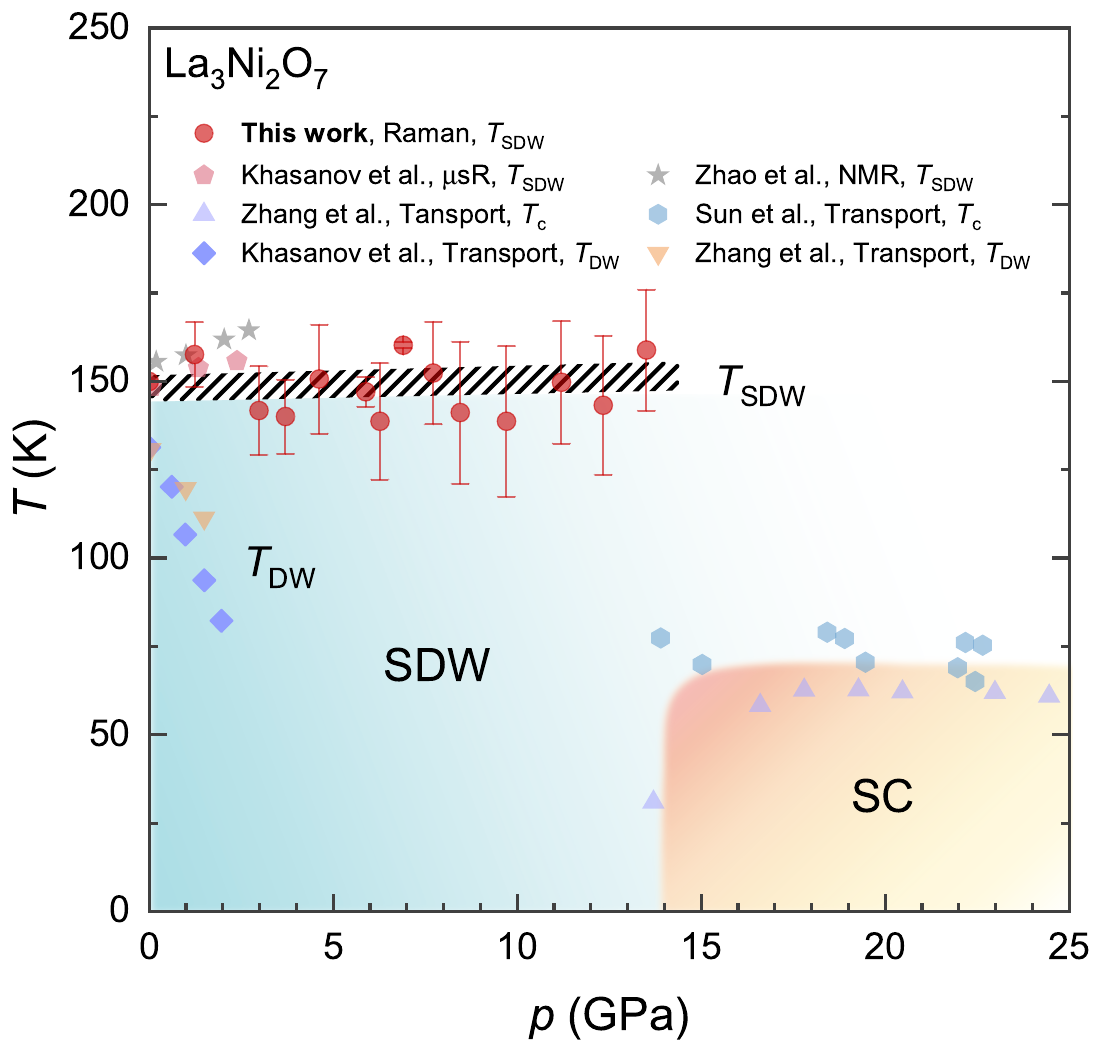}
    \caption{\textbf{Pressure–temperature ($p–T$) phase diagram of \LNO.}  
           The \TSDW\ values are extracted from the temperature evolution of the integrated spectral weight, plotted as red solid circles. Error bars represent the maximum deviation from the averaged \(T_\text{SDW}\) during the four linear-intersection fittings (see Supplemental Materials~\cite{supplement} for more details). The hatched region is the linear fitting of $T_\text{SDW}$. Additional \TSDW\ points at low pressures from $\mu$SR and NMR studies are taken from Refs.~\cite{Khasanov:2025,Zhao2025SB.70.1239}. The purple diamonds and orange triangles mark an additional density-wave transition ($T_{\text{DW}}$)~\cite{Khasanov:2025,Zhang:2024}, whose origin remains under debate. Superconducting $T_{\text{c}}$ values are adopted from Refs.~\cite{Zhang:2024,Sun:2023}. The cyan and orange shaded regions correspond to the SDW and superconducting phases, respectively.}
	\label{fig:phasediagram}
\end{figure}

\begin{figure}[ht!]
  \centering
    \includegraphics[width=8cm]{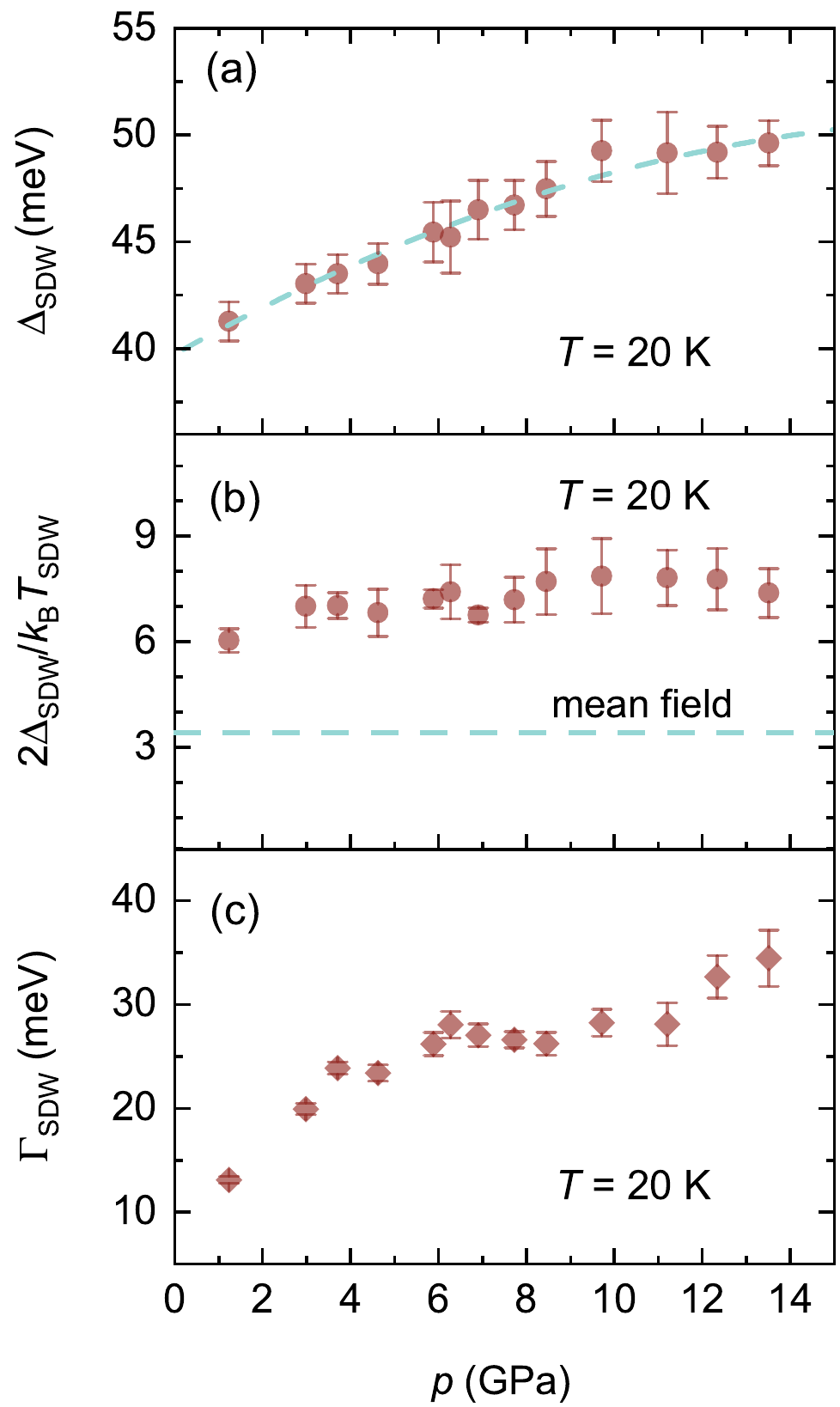}
  \caption{
\textbf{Pressure evolution of SDW-related Raman} parameters. (a) SDW characteristic energy gap $\Delta_{\text{SDW}}$ at $T$ = 20~K as a function of pressure. The dashed line serves as a guide to the eye. (b) Pressure dependence of the coupling ratio $2\Delta_\text{SDW}/k_B T_\text{SDW}$ at $T$ = 20~K, with the mean-field value marked by the dashed line. (c) linewidth $\Gamma_\text{SDW}$ at 20~K. $\Gamma_\text{SDW}$ is defined as the full width at half maximum of the Lorentzian peak. SDW peak position and linewidth error bars represent 95\% confidence intervals derived from fitting procedures.}
	\label{fig:linewidth}
\end{figure}

Combining our electronic Raman measurements with previously reported data~\cite{Khasanov:2025,Zhao2025SB.70.1239,Zhang:2024,Hou2023CPL.40.117302,Sun:2023} including superconducting data from samples of the same source as those in Ref.~\cite{Sun:2023}, we construct the $p$\text{\ensuremath{-}}$T$ phase diagram of \LNO\ (Fig.~\ref{fig:phasediagram}), where the cyan and orange shaded regions denote the SDW-ordered and superconducting regimes, respectively. The SDW transition temperature $T_{\mathrm{SDW}}$, marked by red solid circles, increases gradually with pressure, with previous NMR and $\mu$SR measurements at lower pressures~\cite{Khasanov:2025,Zhao2025SB.70.1239}.We note, however, that the pressure dependence of the Raman-derived $T_{\mathrm{SDW}}$ is weaker than that reported by NMR and $\mu$SR. This difference likely reflects the distinct sensitivities of the probes: NMR and $\mu$SR directly detect the evolution of the magnetic degrees of freedom, whereas Raman spectroscopy senses the SDW transition indirectly through the associated reconstruction of the electronic band structure and electron--hole excitation spectrum. Consequently, the characteristic temperature extracted from Raman reflects the development of the SDW-induced electronic reconstruction and need not exhibit exactly the same pressure dependence as the magnetic transition temperature obtained from direct magnetic probes. Meanwhile, the SDW-related Raman peak broadens progressively with pressure, indicating enhanced damping and a less well-defined SDW-related electronic response (described in Fig.~\ref{fig:linewidth}). Superconductivity emerges near 14~GPa within the SDW-ordered regime. This phase diagram provides the context for the spectroscopic analysis of the pressure evolution of the SDW state presented below.

\begin{figure}[ht!]
  \centering
  \includegraphics[width=8cm]{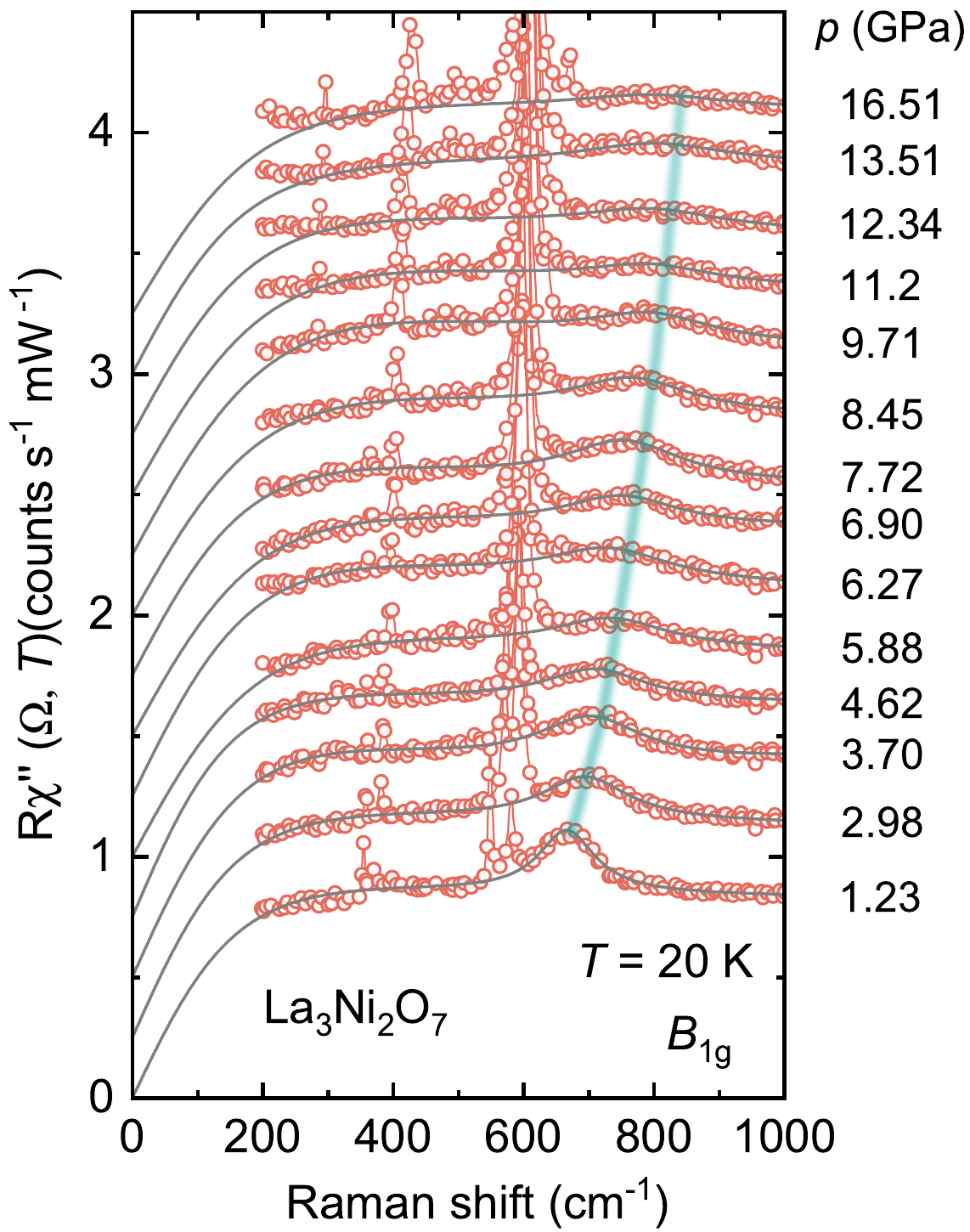}
  \caption{ \textbf{Pressure evolution of the \Blg electronic Raman response.} The red open circles denote the \Blg electronic Raman response measured at 20\,K under various pressures from 1.23 to 16.51\,GPa. Black solid lines represent fits to the experimental spectra. The cyan line serves as a guide to the eye illustrating the evolution of the SDW gap-related peak.}
	\label{fig:energygap}
\end{figure}

%The evolution of magnetic order is tracked through the SDW gap (e.g., at $\sim$ 700\,\wn at 20\,K and 1.23~GPa) in the \Blg spectra as in our previous report~\cite{GeHe:2025}. Figure~\ref{fig:energygap} presents the pressure evolution of the \Blg-symmetric electronic Raman response of \LNO measured at $T$ = 20~K over the range 1.23-16.51 GPa. A broad peak is observed at most measured pressures. This feature originates from interband electronic transitions across the SDW gap ~\cite{Chauviere2010PRB.82.180521,Chauviere2011PRB.84.104508,GeHe:2025,Loret2019NP.15.771}. With increasing pressure, the SDW peak exhibits a monotonic blue shift (highlighted by the cyan guide line) and  progressive broadening, demonstrating a continuous increase in the SDW gap magnitude. 

 The \Blg\ Raman response measured at 20~K and 1.23~GPa exhibits a broad SDW-related feature centered near 700~\wn. This feature has previously been associated with interband electronic transitions across the SDW gap, although a two-magnon contribution has also been proposed~\cite{Gim2026arXiv.2602.05365}. In the following, we adopt the SDW-gap-related interband excitation as our primary working interpretation, while noting that the present Raman data do not completely exclude a magnetic contribution~\cite{Chauviere2010PRB.82.180521,Chauviere2011PRB.84.104508,GeHe:2025,Loret2019NP.15.771}. Figure~\ref{fig:energygap} shows the pressure-dependent \Blg-symmetric Raman response of \LNO\ at 20~K over the range of 1.23--16.51~GPa. The SDW-related peak remains discernible over nearly the entire pressure range. With increasing pressure, its characteristic energy shifts progressively upward, as highlighted by the cyan guide line, indicating a continuous increase in the characteristic SDW-related energy scale. Concurrently, the feature broadens, suggesting reduced coherence of the SDW-related electronic response.

%Systematic temperature- and pressure-dependent Raman measurements enable us to extract both the SDW gap $\Delta_{\text{SDW}}$ and \TSDW\ (see Supplemental Materials~\cite{supplement} for details).  To decompose the electronic Raman response, the low-energy itinerant carrier continuum is described via the memory function method, while the SDW-associated electronic peak is empirically fitted using a Lorentzian function~\cite{GeHe:2025}. The solid black lines in Fig.~\ref{fig:energygap} represent the fitting results, which allow precise and reliable extraction of the peak position and linewidth for the SDW feature at each pressure.

The SDW gap $\Delta_{\mathrm{SDW}}$ and transition temperature $T_{\mathrm{SDW}}$ are extracted from systematic temperature- and pressure-dependent Raman measurements (see Supplemental Material~\cite{supplement} for details). In the following, $\Delta_{\mathrm{SDW}}$, $T_{\mathrm{SDW}}$, and $\Gamma_{\mathrm{SDW}}$ are used as phenomenological parameters characterizing the Raman response associated with the SDW state. Their pressure evolution is expected to reflect the corresponding SDW-related electronic reconstruction probed by Raman spectroscopy. The low-energy electronic continuum is modeled using the memory-function formalism, whereas the SDW-related Raman peak is fitted with a Lorentzian function~\cite{GeHe:2025}. Representative fits are shown as solid black lines in Fig.~\ref{fig:energygap}, from which the peak position and linewidth are obtained at each pressure. Figure~\ref{fig:linewidth} summarizes the pressure dependence of the extracted SDW parameters. The SDW gap $\Delta_{\mathrm{SDW}}$ [Fig.~\ref{fig:linewidth}(a)] increases monotonically with pressure, from approximately 41.3~meV at 1.23~GPa to about 49.6~meV near 13~GPa, corresponding to a $\sim20\%$ enhancement. Notably, $\Delta_{\mathrm{SDW}}$ and $T_{\mathrm{SDW}}$ evolve in parallel with increasing pressure. Consequently, the dimensionless coupling ratio $2\Delta_{\mathrm{SDW}}/(k_{\mathrm B}T_{\mathrm{SDW}})$ remains approximately pressure independent around $\sim7.5$ over the entire pressure range [Fig.~\ref{fig:linewidth}(b)]. This value is well above the weak-coupling mean-field value of 3.52, indicating that the SDW state in \LNO\ resides in the strong-coupling regime. The robust scaling between $\Delta_{\mathrm{SDW}}$ and $T_{\mathrm{SDW}}$ indicates that pressure enhances the characteristic energy scale of the SDW state while preserving its strong-coupling character.

\section{Discussion}

The observed enhancement of the characteristic energy of the SDW-related Raman peak under pressure can be naturally understood in terms of pressure-induced modifications of the Ni-O bonding and the associated magnetic exchange interactions. Recent theoretical studies based on the $t$\text{\ensuremath{-}}$J$ model suggest that the interlayer exchange interaction is substantially stronger than its intralayer counterpart and plays an important role in the emergence of superconductivity in bilayer nickelates~\cite{Wucongjun:2024,Wu2024SCPMA,Tian2025PRB.112.014520,Bejas2025PRB.111.144514,Qin2026arXiv.2606.15298}. Within the strong-coupling limit, the Hund's coupling between the Ni $d_{z^2}$ and $d_{x^2-y^2}$ orbitals allows the low-energy physics of \LNO to be described by an effective single-orbital bilayer $t$--$J$ model~\cite{Wucongjun:2024}. In this framework, the superexchange interaction scales approximately as $J\propto4t^2/U$, where $t$ is the hopping integral and $U$ is the on-site Coulomb repulsion. Hydrostatic compression shortens the Ni-O bond length and modifies the Ni-O-Ni bond angle, thereby enhancing the overlap between the Ni $3d_{z^2}$ and O $2p_z$ orbitals and increasing the effective hopping integral $t$~\cite{Sun:2023,Verraes2025arXiv.2502.19501}. Recent multiorbital calculations also suggest that the pressure-induced increase in crystal-field splitting may further enhance $T_{\mathrm{SDW}}$ by driving the $d_{z^2}$ orbital toward half filling and strengthening its contribution to magnetic order~\cite{IlyaCDWPRB:2026}. The resulting enhancement of magnetic exchange interactions provides a natural explanation for the observed increases in both $\Delta_{\mathrm{SDW}}$ and $T_{\mathrm{SDW}}$. Since superconductivity emerges in the same pressure regime~\cite{Zhang:2024,Hou2023CPL.40.117302,Sun:2023}, the pressure-tuned magnetic interactions may also be relevant for understanding the relationship between magnetism and superconductivity in bilayer nickelates.

The pressure dependence of the SDW-related Raman peak linewidth, $\Gamma_{\mathrm{SDW}}$, provides complementary information on the evolution of the SDW-related electronic response. As shown in Fig.~\ref{fig:linewidth}(c), $\Gamma_{\mathrm{SDW}}$ increases progressively with pressure. Because the Raman feature arises from interband electronic transitions between SDW-reconstructed states, its linewidth is sensitive to both the distribution of the corresponding transition energies and the electronic scattering rate~\cite{Devereaux:1994,Devereaux:2007}. The increase in $\Gamma_{\mathrm{SDW}}$ therefore suggests enhanced damping and/or a broader distribution of these transitions, indicating that the SDW-related electronic response becomes progressively less well defined under pressure.

%\rev{Pressure-induced spatial inhomogeneity in the SDW state may provide an additional source of linewidth broadening. First-order phase transitions can proceed through nucleation and phase coexistence, giving rise to spatially inhomogeneous domains near the transition~\cite{Binder1987}. Nanoscale phase coexistence has been directly observed in several strongly correlated systems~\cite{Qazilbash2007,Uehara1999}, while short-range and fluctuating DW correlations are well established in cuprate superconductors~\cite{Boschini2021,Lee2021NP,Lu2022PRB}. In \LNO, recent transport measurements associate the emergence of superconductivity with a first-order transition~\cite{Shi2025NP.21.1780}, raising the possibility of spatial coexistence between short-ranged correlated SDW and superconducting regions. Such inhomogeneity could broaden the distribution of SDW-related excitation energies and thereby contribute to the increase in $\Gamma_{\mathrm{SDW}}$.}

Pressure-induced spatial inhomogeneity may provide an additional source of linewidth broadening. In systems with competing electronic orders, spatial phase coexistence and inhomogeneous domains can develop near a phase boundary, particularly when the transition is sharp or weakly first order~\cite{Binder1987}. Such nanoscale phase coexistence has been directly observed in several strongly correlated systems~\cite{Qazilbash2007,Uehara1999}, while short-range and fluctuating DW correlations are well established in the pseudogap phase of cuprate superconductors~\cite{Boschini2021,Lee2021NP,Lu2022PRB}. In bilayer nickelates, superconductivity emerges as the SDW state is suppressed at high pressure, suggesting competition between the two orders~\cite{Wang:2024,Sun:2023,Zhang:2024,WangbulkSC:2024,Zhu:2024,Zhangjunjie:2025,Shi2025NP.21.1780}. This competition may favor spatially inhomogeneous regions with different relative strengths of SDW and superconducting correlations near the phase boundary\cite{Mandyam2026}. Such inhomogeneity could broaden the distribution of SDW-related excitation energies and thereby contribute to the increase in $\Gamma_{\mathrm{SDW}}$.

Together with the simultaneous increases in $\Delta_{\mathrm{SDW}}$ and $T_{\mathrm{SDW}}$, these results reveal an unconventional pressure evolution: the characteristic energy of the SDW-related Raman peak is enhanced, whereas the SDW-related electronic response becomes increasingly broadened. The onset of superconductivity in the same pressure regime further suggests a possible interplay between SDW order and superconductivity, consistent with theoretical proposals that pressure weakens long-range SDW order while enhancing magnetic fluctuations~\cite{Le2025arXiv.2501.14665v3,Botzel2026arXiv.2606.23022v1}.

The pressure evolution of magnetic correlations has been extensively investigated in cuprates and iron-based superconductors. For cuprates, in La-based compounds near the $1/8$ hole doping, spin-stripe order exhibits a strong interplay with superconductivity under pressure or uniaxial stress~\cite{Guguchia2013NJP.15.093005,Guguchia2020PRL.125.097005,Kamminga2023PRB.107.144506,Guguchia2024PNAS}. In iron-based superconductors, hydrostatic pressure generally suppresses SDW order, as widely observed in the 122 family~\cite{Torikachvili2008PRB.78.104527,Ishikawa2009PRB.79.172506,Colombier2009PRB.79.224518,Alireza2008JPCM.21.012208,Ahilan2008JPCM.20.472201,Uykur2015PRB.92.245133}. FeSe displays a more complex, nonmonotonic evolution, in which pressure initially stabilizes magnetic order before eventually suppressing it as superconductivity develops~\cite{Sun2016NC.7.12146}. By contrast, both $T_{\mathrm{SDW}}$ and $\Delta_{\text{SDW}}$ in \LNO\ increase monotonically with pressure, with no downturn observed over the measured pressure range. This contrasting behavior suggests a distinct pressure response of magnetic order in bilayer nickelates.

In summary, we reveal an unconventional pressure evolution of SDW order in~\LNO. The SDW energy scale increases continuously with pressure, while the Raman response becomes progressively broadened, indicating enhanced damping of the associated electronic excitation. The pressure-independent coupling ratio demonstrates that the strong-coupling nature of the SDW state is preserved, providing new constraints on the relationship between magnetism and superconductivity in bilayer nickelates.\\

\textit{Acknowledgements}\text{\ensuremath{-}}We thank Fuchun Zhang, Zhengcheng Gu and Yilin Wang for fruitful discussions. We thank the staff members of the Infrared spectroscopy and microspectroscopy Beamline (https://cstr.cn/31131.02.HLS.IRSM) at the Hefei Light Source (https://cstr.cn/31131.02.HLS) for providing technical support and assistance in data collection and analysis. This work was supported by the New Cornerstone Science Foundation (Grant No. NCI202211 (D.L.F.)), and Quantum Science and Technology-National Science and Technology Major Project (Grant No. 2021ZD0302803(D.L.F.)), and HFNL Self-Deployed Project (Grant No. ZB2602000303 (Z.D.)), and the National Natural Science Foundation of China (Grant No. 12474473(G.H.)). Work at SYSU was supported by the National Natural Science Foundation of China (Grant No. 12425404, U25A20193, 92565303), the Fundamental and Interdisciplinary Disciplines Breakthrough Plan of the Ministry of Education of China (JYB2025XDXM403), the Guangdong Major Project of Basic Research (2025B0303000004), the Guangdong Provincial Key Laboratory of Magnetoelectric Physics and Devices (Grant No. 2022B1212010008), and Research Center for Magnetoelectric Physics of Guangdong Province (Grant No. 2024B0303390001). \\

%\noindent\textbf{Author contributions}\\
%XZ, ZQ, MW and DLF conceived the project. XZ, RL, SX, LQ, HL, CH and HZ performed the Raman measurements. XZ, SX and JS analysed the Raman data. MH, DH, and MW synthesised and characterised the samples. XZ, GH and ZD wrote the manuscript with comments from all authors.\\

%\noindent\textbf{Competing interests}\\
%The authors declare no competing interests.

\bibliography{refs}

\end{document}